# A Faithful Deep Sensitivity Estimation for Accelerated Magnetic Resonance Imaging

Zi Wang, Haoming Fang, Chen Qian, Boxuan Shi, Lijun Bao, Liuhong Zhu, Jianjun Zhou, Wenping Wei, Jianzhong Lin, Di Guo, and Xiaobo Qu

*Abstract*—Magnetic resonance imaging (MRI) is an essential diagnostic tool that suffers from prolonged scan time. To alleviate this limitation, advanced fast MRI technology attracts extensive research interests. Recent deep learning has shown its great potential in improving image quality and reconstruction speed. Faithful coil sensitivity estimation is vital for MRI reconstruction. However, most deep learning methods still rely on pre-estimated sensitivity maps and ignore their inaccuracy, resulting in the significant quality degradation of reconstructed images. In this work, we propose a Joint Deep Sensitivity estimation and Image reconstruction network, called JDSI. During the image artifacts removal, it gradually provides more faithful sensitivity maps with high-frequency information, leading to improved image reconstructions. To understand the behavior of the network, the mutual promotion of sensitivity estimation and image reconstruction is revealed through the visualization of network intermediate results. Results on *in vivo* datasets and radiologist reader study demonstrate that, for both calibration-based and calibrationless reconstruction, the proposed JDSI achieves the state-of-the-art performance visually and quantitatively, especially when the acceleration factor is high. Additionally, JDSI owns nice robustness to patients and autocalibration signals.

This work was supported in part by the National Natural Science Foundation of China under grants 62122064, 61971361, 62331021, 62371410, and 62071405, the Natural Science Foundation of Fujian Province of China under grant 2023J02005 and 2021J011184, the President Fund of Xiamen University under grant 20720220063, the Xiamen University Nanqiang Outstanding Talents Program, and the China Scholarship Council under Grant 202306310177. *(Zi Wang and Haoming Fang contributed equally to this work.) (Corresponding author: Xiaobo Qu, email: quxiaobo@xmu.edu.cn)*

Zi Wang, Haoming Fang, Chen Qian, Boxuan Shi, Lijun Bao, and Xiaobo Qu are with the Department of Electronic Science, Biomedical Intelligent Cloud R&D Center, Fujian Provincial Key Laboratory of Plasma and Magnetic Resonance, National Institute for Data Science in Health and Medicine, Xiamen University, Xiamen 361005, China.
Liuhong Zhu and Jianjun Zhou are with the Department of Radiology, Zhongshan Hospital (Xiamen), Fudan University, Xiamen 361006, China.
Wenping Wei is with the Department of Radiology, The First Affiliated Hospital of Xiamen University, Xiamen 361005, China.
Jianzhong Lin is with the Department of Radiology, Zhongshan Hospital affiliated to Xiamen University, Xiamen 361005, China.
Di Guo is with the School of Computer and Information Engineering, Xiamen University of Technology, Xiamen 361024, China.

## I. INTRODUCTION

MAGNETIC resonance imaging (MRI) is a leading non-invasive and non-radioactive diagnostic modality in modern medical science [1]. However, MRI scans usually require a long acquisition time, and thus fast MRI technology attracts extensive research interests [2]. Parallel imaging [3, 4] is a widely used multi-coil acceleration technique in clinical systems, while sparse sampling [2, 5] provides a complementary scheme for further obtaining a higher acceleration factor (AF).

Over the past two decades, coil correlation modeling has always been a key issue in advanced MRI reconstructions, which can be categorized into explicit and implicit sensitivity estimations. The former typically pre-acquires [3] or pre-estimates (e.g., using ESPIRiT [6]) the sensitivity maps from the autocalibration signal (ACS), and further applies sparse regularizations [2, 7-11] to coil-combined images to reduce the reconstruction error. Moreover, sensitivity maps could also be regularized using polynomial functions (JSENSE [12]), so as to continuously update the sensitivity estimation with the image reconstruction. For implicit methods, two main strategies include the k-space kernel learning (e.g., GRAPPA [4] and SPIRiT [13]) and structured low-rank (e.g., SAKE [14] and P-LORAKS [15]), which can be applied efficiently in the calibrationless reconstruction. Recently, based on their good reconstruction performance and complementarity, the two approaches have been combined for further improvement [16, 17].

At present, deep learning has shown its great potential in fast MRI to improve image quality and reconstruction speed [18-21]. Like conventional methods, there are also implicit and explicit strategies for deep learning when dealing with multi-coil data. Several implicit methods without sensitivity map estimation utilize multiple convolutional layers to achieve k-space interpolation (e.g., DOTA [22], HDSLR [23], and ODLS [24]) and is robust to the number of ACS lines. Whereas most existing deep learning methods like MoDL [25] involve explicit sensitivity maps pre-estimated using ESPIRiT [6], and they can work efficiently under the moderate AF [26-29].

When the AF is high or ACS lines are limited, the conventional sensitivity estimation will be inaccurate, and the error may affect the image reconstruction and bring significant image degradation [12, 14]. To overcome this problem, one interesting attempt is to estimate sensitivity maps through a deep network (DeepSENSE [30]) but its reconstruction performance is limited by the subsequent SENSE reconstruction. To avoid pre-estimating, another work employs a U-Net as the sensitivity estimator and inserts it at the beginning of the network (E2E-VarNet [31]), but ignores the importance of sensitivity refinements. To solve this problem, some joint image and sensitivity reconstruction networks are proposed to update sensitivity maps with conventional linear operations (Joint-ICNet [32]) or compact kernels (Deep-JSENSE [33]). Their promising results arouse our interest, however, they are hard to obtain faithful sensitivity maps since they only use the ACS region to get low-resolution sensitivity maps and lose the abundant high-frequency information. Thus, we found that they still need to be improved to achieve better network interpretability (Section III.C), lower reconstruction error (Section IV.C), and greater robustness to patients (Section IV.D).

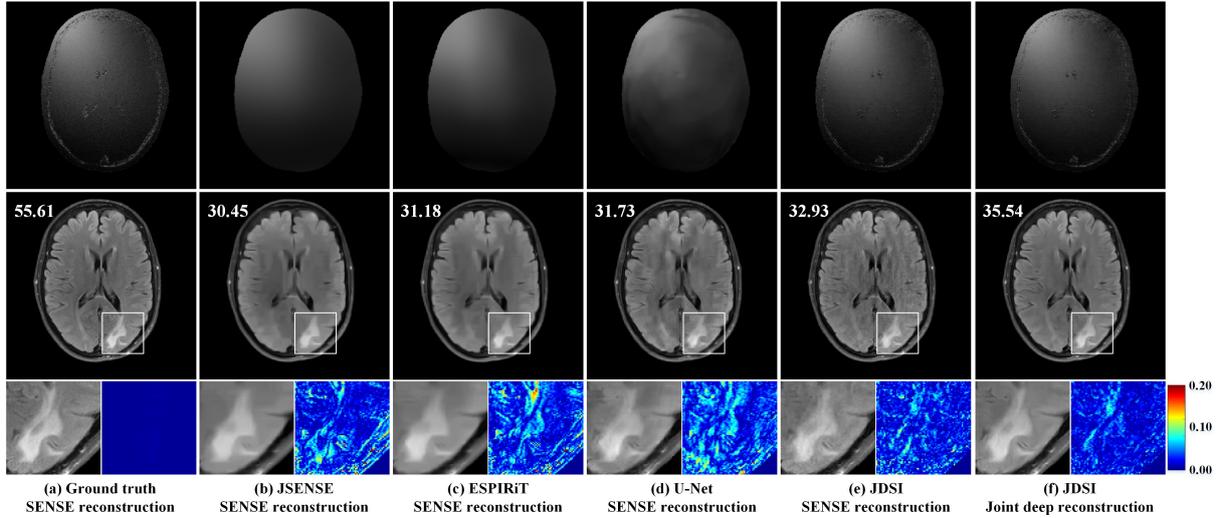

**Fig. 1.** Patient brain reconstructions under a 1D Cartesian undersampling pattern with AF=6. First row: (a) Ground truth of 1$^{st}$ coil which is generated via dividing the individual fully sampled coil image by the SoS image of fully sampled multi-coil images. Estimated sensitivity maps using (b) JSENSE [12], (c) ESPIRiT [6], (d) U-Net [31], and (e)-(f) the proposed method. Second row: (a)-(e) Reconstructed images using SENSE-based reconstruction [11] and (f) the proposed joint deep reconstruction. Third row: (a)-(f) Magnified lesions and the corresponding error maps. Note: PSNR(dB) is listed for each reconstruction.

A faithful sensitivity estimation is important for high-quality MRI reconstruction using explicit methods [6, 12, 30]. In this paper, the ground truth of coil sensitivity maps is obtained via dividing the individual fully sampled coil image by the sum of squares (SoS) of fully sampled multi-coil images. With these ground truth sensitivity maps with high-frequency information, Fig. 1(a) illustrates that an almost error-free lesion reconstruction can be achieved. In contrast, the estimated sensitivity maps which are much closer to the ground truth have better SENSE-based reconstructions [11] (Figs. 1(b)-(e)). And our proposed joint deep learning can further suppress artifacts and reconstruct the lesion margin clearly (Fig. 1(f)). This example clearly shows that a better sensitivity map with high-frequency information leads to a lower reconstruction error and better lesion details preservation, no matter for using the conventional SENSE or deep learning reconstructions.

Our joint deep network, called JDSI, is proposed to simultaneously achieve the faithful estimation of sensitivity maps and reconstruction of high-quality images. The main contributions are summarized as follows:

1) The designed network estimates sensitivity maps using all measured k-space and explicitly constrains these maps in the loss function. It provides faithful sensitivity maps with high-frequency information.

2) We explain our network by visualizing the intermediate sensitivity maps and images. It is found that the alternating sensitivity estimation and image artifacts removal promote each other, leading to high-quality reconstructions.

3) Extensive results on *in vivo* brain datasets show that, for both calibration-based and calibrationless reconstruction, the proposed JDSI provides superior reconstruction performance.

4) The reader study shows the overall image quality of JDSI steps into the excellent level under three experienced radiologists' evaluations, indicating its reliability in clinical diagnosis even under highly accelerated scanning (AF=6).

## II. Problem Formulation

Magnetic resonance image reconstruction problem is to recover a desired image $\mathbf{x} \in \mathbb{C}^N$ from the measured multi-coil k-space $\mathbf{y} \in \mathbb{C}^{JN}$ which is undersampled by an operator $\mathcal{U}$ with non-acquired positions zero-filled. The forward model can be defined as

$$\mathbf{y} = \mathcal{U}\mathcal{F}\mathbf{S}\mathbf{x} + \boldsymbol{\varepsilon}, \tag{1}$$

where $\mathbf{S} = [\mathbf{S}_1;...;\mathbf{S}_j;...;\mathbf{S}_J] \in \mathbb{C}^{JN \times N}$ is the set of sensitivity maps. $N$ represents the dimension of the vectorized image and $J$ is the number of coils. $\mathbf{S}_j \in \mathbb{C}^{N \times N}$ is a diagonal matrix which denotes the sensitivity map of the $j^{\text{th}}$ coil. $\mathcal{F}$ is the Fourier transform, $\boldsymbol{\varepsilon} \in \mathbb{C}^{JN}$ is the additive Gaussian noise.

For this ill-posed problem in Eq. (1), assuming that the sensitivity maps $\mathbf{S}$ are pre-acquired or pre-calculated from the ACS region, it can be solved using the following optimization model with a regularization term $\mathcal{R}_1(\cdot)$, e.g., the $l_1$-norm in image sparsity [2, 7-11]:

$$\min_{\mathbf{x}} \frac{1}{2} \|\mathbf{y} - \mathcal{U}\mathcal{F}\mathbf{S}\mathbf{x}\|_2^2 + \mathcal{R}_1(\mathbf{x}). \tag{2}$$

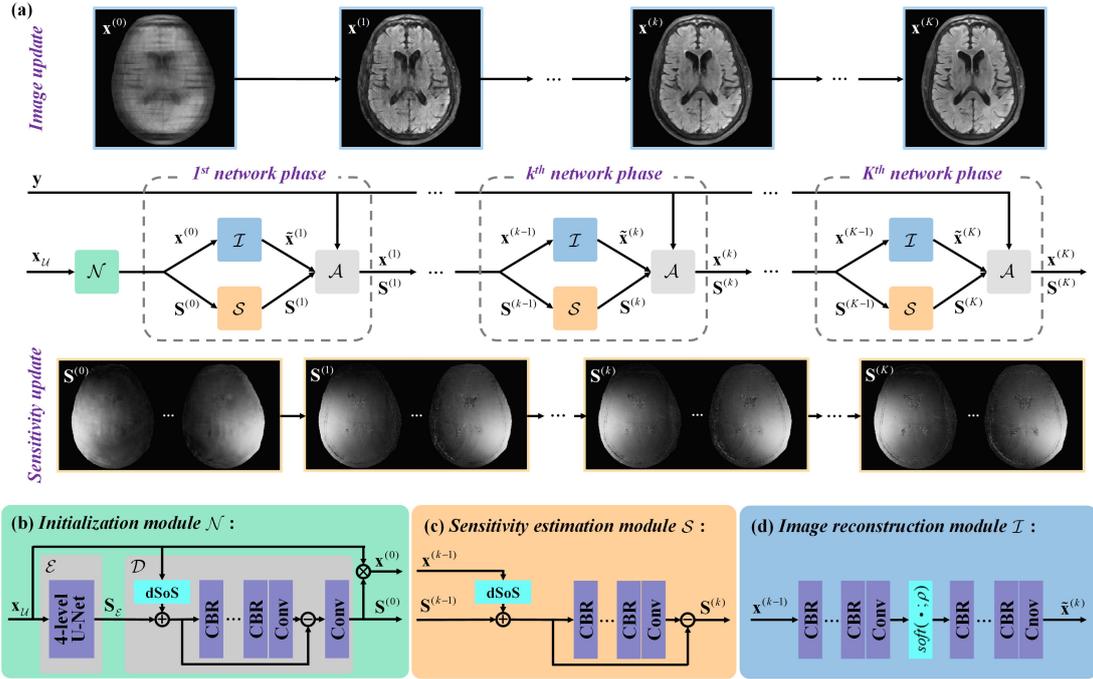

**Fig. 2.** The proposed JDSI for accelerated MRI. (a) The recursive JDSI architecture includes (b) initialization module $\mathcal{N}$, (c) sensitivity estimation module $\mathcal{S}$, (d) image reconstruction module $\mathcal{I}$, and data consistency module $\mathcal{A}$. Note: "dSoS" is the operation that is divided by the sum of squares images. "CBR" is the convolution, batch normalization, and ReLU. "Conv" is the convolution.

However, the pre-estimated sensitivity maps may not be accurate due to inconsistent pre-scan or limited ACS lines [12]. Then, inaccuracy sensitivity estimation will affect the image reconstruction and result in artifacts [23, 30].

To improve the sensitivity estimation, a joint sensitivity estimation and image reconstruction model is proposed [12]. It simultaneously updates the sensitivity maps and image. After introducing another regularization term of the sensitivity maps $\mathcal{R}_2(\cdot)$, e.g., polynomial regularization [12], Eq. (2) is extended as:

$$\min_{\mathbf{x},\mathbf{S}} \frac{1}{2}\|\mathbf{y}-\mathcal{U}\mathcal{F}\mathbf{S}\mathbf{x}\|_2^2 + \mathcal{R}_1(\mathbf{x}) + \mathcal{R}_2(\mathbf{S}). \quad (3)$$

To solve the optimization problem in Eq. (3), the conventional reconstruction methods update $\mathbf{S}$ and $\mathbf{x}$ alternatively, until the convergence [12]. The reconstruction process is time-consuming due to the usage of an iterative optimization algorithm, while the regularization constraints and their parameters need to be manually determined.

### III. PROPOSED METHOD

In this section, we propose a joint deep network whose architecture is inspired by the alternating iterative idea of the conventional reconstruction. Then, the detailed network implementation is presented.

#### A. JDSI: Joint Deep Sensitivity Estimation and Image Reconstruction Network

Here, we propose a Joint Deep Sensitivity estimation and Image reconstruction network (JDSI) to alternate between updating $\mathbf{S}$ and $\mathbf{x}$, using a deep learning scheme. Their updates correspond to two network modules, i.e., the sensitivity estimation module $\mathcal{S}$ and image reconstruction module $\mathcal{I}$. An initialization module is also integrated at the network beginning to provide proper initial image and sensitivity maps. Besides, to ensure the network reconstructions are aligned with the acquired data, the data consistency module is added. It is noted that all sensitivity maps are normalized to satisfy $\sum_{j=1}^{J} \mathbf{S}_j^* \mathbf{S}_j = \mathbf{I}$, where $\mathbf{I}$ is an identity matrix and the superscript $*$ represents the adjoint operation.

Fig. 2 gives the network architecture of our proposed JDSI. Specifically, our network firstly estimates initial sensitivity maps from the undersampled multi-coil k-space and provides the initial coil-combined image. Second, the sensitivity estimation module is utilized to refine sensitivity maps from the whole multi-coil k-space. Third, the undersampling artifacts are removed through the image reconstruction module. Fourth, reconstructed images and sensitivity maps are projected back to the k-space and then forced to maintain the data consistency to the measured k-space. After repeating the last three modules $K$ times, the final reconstructed image can be obtained.

*1) Initialization module*

Fig. 2(b) shows the initialization module that aims to initialize sensitivity maps and the coil-combined image. We employ two sub-networks (i.e., $\mathcal{E}, \mathcal{D}$), to obtain initial sensitivity maps $\mathbf{S}^{(0)}$ from undersampled multi-coil image $\mathbf{x}_\mathcal{U} = \mathcal{F}^*\mathcal{U}^*\mathbf{y}$. The overall process is written as

$$\mathbf{S}^{(0)} = \mathcal{N}(\mathbf{x}_\mathcal{U}; \boldsymbol{\theta}_\mathcal{N}), \quad (4)$$

where $\boldsymbol{\theta}_\mathcal{N}$ is the network parameters for the mapping $\mathcal{N}$. The superscript $*$ represents the adjoint operation. Two sub-networks are defined as

$$\begin{cases} \mathbf{S}_\mathcal{E} = \mathcal{E}(\mathbf{x}_\mathcal{U}; \boldsymbol{\theta}_\mathcal{E}) \\ \mathbf{S}^{(0)} = \mathcal{D}(\mathbf{S}_\mathcal{E}, \mathbf{x}_\mathcal{U}; \boldsymbol{\theta}_\mathcal{D}) \end{cases}, \quad (5)$$

where $\boldsymbol{\theta}_\mathcal{E}, \boldsymbol{\theta}_\mathcal{D}$ are the network parameters for the mapping $\mathcal{E}, \mathcal{D}$, respectively. Through an encoder-decoder network $\mathcal{E}$, the undersampled image $\mathbf{x}_\mathcal{U}$ is obtained with just low-resolution information, then sensitivity maps $\mathbf{S}_\mathcal{E}$ are generated via dividing the individual coil images by the SoS image [31]. Unlike some existing methods [6, 30] that only use the ACS region to get low-resolution sensitivity maps, we exploit the whole acquired k-space [12] to capture more high-frequency information. After that, a denoiser $\mathcal{D}$ is designed to remove the residual noises and artifacts, then the initial sensitivity maps $\mathbf{S}^{(0)}$ are obtained.

Once the initial sensitivity maps $\mathbf{S}^{(0)}$ are generated, the initial coil-combined image can be obtained as $\mathbf{x}^{(0)} = (\mathbf{S}^{(0)})^* \mathbf{x}_\mathcal{U}$, which commonly has strong artifacts.

*2) Sensitivity estimation module*

Fig. 2(c) shows the sensitivity estimation module that corresponds to update $\mathbf{S}$, which inputs are the fusion of $\mathbf{x}^{(k-1)}$ and $\mathbf{S}^{(k-1)}$:

$$\mathbf{S}^{(k)} = \mathcal{S}(\mathbf{S}^{(k-1)}, \mathbf{x}^{(k-1)}; \boldsymbol{\theta}_\mathcal{S}), \quad (6)$$

where $k$ denotes the number of network phase, and $\boldsymbol{\theta}_\mathcal{S}$ is the network parameters for the mapping $\mathcal{S}$. The mapping is a multi-layer convolutional network that refines the sensitivity maps estimation using the whole k-space rather than only the ACS region.

In this way, more high-frequency information can be retained to further approximate the ground truth sensitivity maps. It prevents the errors of the estimated sensitivities from propagating to the final reconstruction.

*3) Image reconstruction module*

Fig. 2(d) shows the image reconstruction module that corresponds to update $\mathbf{x}$. Since this module is designed for artifacts removal, here, we utilize the efficient deep thresholding network [24, 27, 34, 35]:

$$\tilde{\mathbf{x}}^{(k)} = \mathcal{I}(\mathbf{x}^{(k-1)}; \boldsymbol{\theta}_\mathcal{I}), \quad (7)$$

where $\boldsymbol{\theta}_\mathcal{I}$ is the network parameters for the mapping $\mathcal{I}$. The mapping involves two multi-layer convolutional networks, and an element-wise soft-thresholding $soft(x; \rho) = \max\{|x|-\rho\} \cdot x/|x|$ is inserted between them, where $\rho$ is the learnable threshold that is initialized to 0.001 and varies at each network phase.

*4) Data consistency module*

In this module, each final output is forced to maintain the data consistency to the measured k-space:

$$\mathbf{x}^{(k)} = \mathcal{A}(\tilde{\mathbf{x}}^{(k)}, \mathbf{S}^{(k)}, \mathbf{y}; \lambda), \quad (8)$$

where $\tilde{\mathbf{x}}^{(k)}$ and $\mathbf{S}^{(k)}$ are the output of the image reconstruction and sensitivity estimation module, respectively. $\lambda$ is a learnable trade-off parameter initialized as $10^6$. Specifically, it is defined as

$$(\mathcal{F}\mathbf{S}^{(k)}\mathbf{x}^{(k)})_n = \begin{cases} (\mathcal{F}\mathbf{S}^{(k)}\tilde{\mathbf{x}}^{(k)})_n, & n \notin \Omega \\ \left(\dfrac{\mathcal{F}\mathbf{S}^{(k)}\tilde{\mathbf{x}}^{(k)} + \lambda \mathbf{y}}{1+\lambda}\right)_n, & n \in \Omega \end{cases}, \quad (9)$$

where $n$ is the index and $\Omega$ is the set of the sampled positions in k-space. Eq. (9) implies that, at the sampled positions, the points should maintain a trade-off with $\mathbf{y}$, while the update of the unsampled points depends entirely on the reconstruction results of the network.

### B. Network Architecture and Loss Function

As shown in Fig. 2, the proposed JDSI is an unrolled recursive network, where the number of total network phase $K$ is set as 5 to balance the reconstruction performance and time/memory consumption (Fig. 3(a)).

Our network architecture is as follows: (1) Initialization module has two sub-networks. The encoder-decoder network $\mathcal{E}$ is a modified 4-level U-Net [36] and the number of filters grows from 32 to a maximum of 256. Denoiser $\mathcal{D}$ consists of 15 convolutional layers, and each layer contains 32 filters, followed by batch normalization and ReLU. (2) Sensitivity estimation module $\mathcal{S}$ is a 5-layer convolutional denoiser, and each layer contains 64 filters, followed by batch normalization and ReLU. (3) Image reconstruction module $\mathcal{I}$ involves an element-wise soft-thresholding and two 4-layer convolutional networks. Each layer contains 32 filters, followed by batch normalization and ReLU. Besides, the size of all convolutional filters in our network is 3×3, and the input and output channels of all sub-networks are adjusted according to the number of data coils. Fig. 3(b)-(g) show that our choices are efficient and they provide good reconstruction performance.

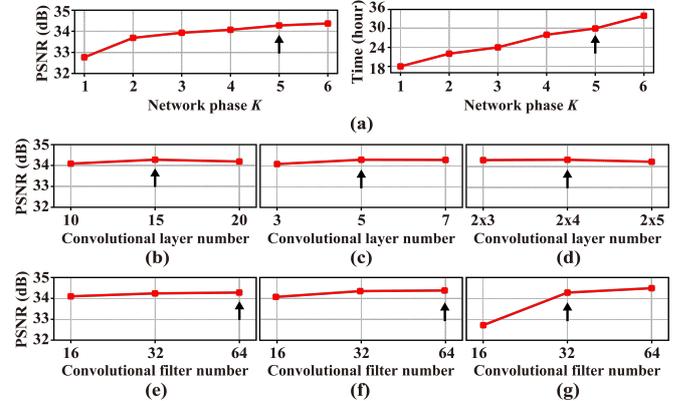

**Fig. 3.** Quantitative comparison of T2 FLAIR brain reconstructions using different network configurations. (a) is the PSNR and training time from different number of network phase. When network phase is set to 6, it consumes the maximum capacity of our GPU (16 GB memory). (b)-(d) (or (e)-(g)) are results from different number of convolutional layers (or filters) in the initialization module (denoiser part), sensitivity estimation module, and image reconstruction module, respectively. Note: The black arrow denotes the chosen setting in the paper. The 1D Cartesian undersampling pattern with AF=8 is used. The means are computed over all test data.

The network weights are Xavier initialized and trained for 200 epochs with the Adam optimizer. Their initial learning rate is set to 0.001 with an exponential decay of 0.99. The batch size is 2. The loss function of the proposed JDSI is defined as:

$$\mathcal{L} = (\mathcal{L}_{coil} + \gamma \mathcal{L}_{combine}) + \alpha \mathcal{L}_{sensitivity} = \mathcal{L}_{image} + \alpha \mathcal{L}_{sensitivity}$$

$$= \frac{1}{KT}\sum_{k=1}^{K}\sum_{t=1}^{T}\left(\begin{array}{l}\left(\left\|\mathbf{x}_t^{ref} - \mathbf{S}_t^{(K)}\mathbf{x}_t^{(K)}\right\|_2^2 + \gamma\left\|(\mathbf{S}_t^{ref})^*\mathbf{x}_t^{ref} - \mathbf{x}_t^{(K)}\right\|_2^2\right) \\ + \alpha\left\|\mathbf{S}_t^{ref} - \mathbf{S}_t^{(K)}\right\|_2^2\end{array}\right), \quad (10)$$

where $T$ is the number of training samples. $\gamma, \alpha$ are set as 0.1 and 0.1, respectively. $\mathcal{L}_{coil}, \mathcal{L}_{combine}, \mathcal{L}_{sensitivity}$ are the multi-coil image loss, combined-coil image loss, and sensitivity map loss, respectively. The former two can be further defined as image loss $\mathcal{L}_{image}$. $\mathbf{x}_t^{ref}$ is the fully sampled multi-coil image of the $t^{th}$ training sample. $\mathbf{S}_t^{ref}$ denote the ground truth sensitivity maps.

The proposed network was implemented in Tensorflow 2.2.0 on a server equipped with dual Intel Xeon Silver 4210 CPUs, 128 GB RAM, and the Nvidia Tesla T4 GPU (16 GB memory). The typical training of JDSI took about 30 hours.

*C. Network Interpretability*

To demonstrate the network interpretability of JDSI, we visualized the intermediate sensitivity maps and the corresponding reconstructed images in Fig. 4.

With the increase of the network phase, the alternating sensitivity estimation and image de-aliasing promote each other. The sensitivity maps gradually approximate the ground truth, while image artifacts are gradually removed, leading to a higher PSNR. This enables JDSI to obtain faithful sensitivity maps and benefits high-quality image reconstruction.

In summary, these observations reveal the mutual promotion of sensitivity estimation and image reconstruction. Besides, like iterative methods, JDSI gradually removes image artifacts with the increase of network phases. Different from the pure end-to-end deep learning methods, it provides a good interpretation to understand the behavior of the network.

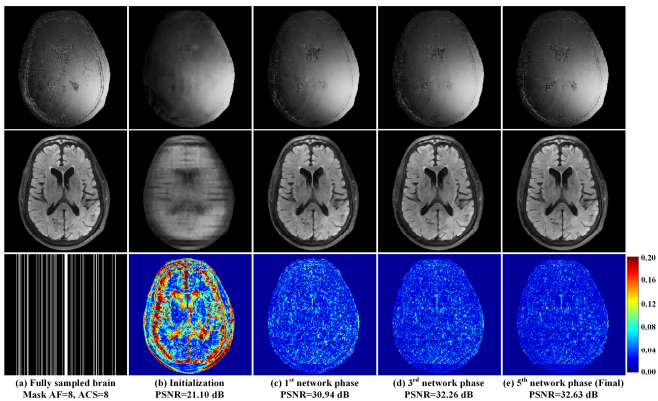

**Fig. 4.** Visualization of the reconstructed images and sensitivity maps at the representative network phase. (a) is the ground truth sensitivity map of $11^{th}$ coil, fully sampled image, and the corresponding undersampling pattern. (b)-(e) are estimated sensitivity maps, reconstructed images, and reconstruction errors of different network phases, from top to bottom. Note: PSNR(dB) is listed for each reconstruction.

## IV. EXPERIMENTAL RESULTS

*A. Datasets and Evaluation Criteria*

In our paper, T2 FLAIR and T2 weighted brain datasets from fastMRI database [37] were used. They were also widely used in many existing deep learning [31, 32], due to the importance of these two contrasts in clinical diagnosis. Specifically, T2 weighted is commonly used to observe tissue lesions, while T2 FLAIR suppresses the cerebrospinal fluid to more clearly display the lesions nearby [38].

For the T2 FLAIR dataset, we randomly used 32 subjects, where each subject contained about 16 slices of size 320×320 and coil 16. 26 subjects were used for training and the remaining for test. For the T2 weighted dataset, we randomly used 37 subjects, where each subject contained about 20 slices of size 320×320 and coil 16. 30 subjects were used for training and the remaining for test. All datasets were fully sampled, and they were retrospectively undersampled for training and test. The number of our method used training and test data was equal to all compared deep learning methods.

To quantitatively evaluate the reconstruction performance, we utilized three evaluation criteria: The relative $l_2$ norm error (RLNE) [9], peak signal-to-noise ratio (PSNR), and structural similarity index (SSIM) [39]. The lower RLNE, higher PSNR, and higher SSIM indicate lower reconstruction error, less image distortions, and better details preservation, respectively.

*B. Ablation Studies*

*1) Initialization module*

It should be especially prudent for a reconstruction method that involves sensitivity maps, and thus requires a reliable estimation as the initial input. Here, we evaluated the importance of the proposed initialization module as mentioned in Eq. (5). We compared the proposed method (Full module) with other two schemes: No module and partial module. The former means that sensitivity maps are initialized via dividing the individual zero-filling coil image by the SoS of zero-filling multi-coil images. The latter represents that sensitivity maps are initialized via only $\mathcal{E}$ in Eq. (5). The network architectures of these two are consistent with JDSI, except for the initialization module.

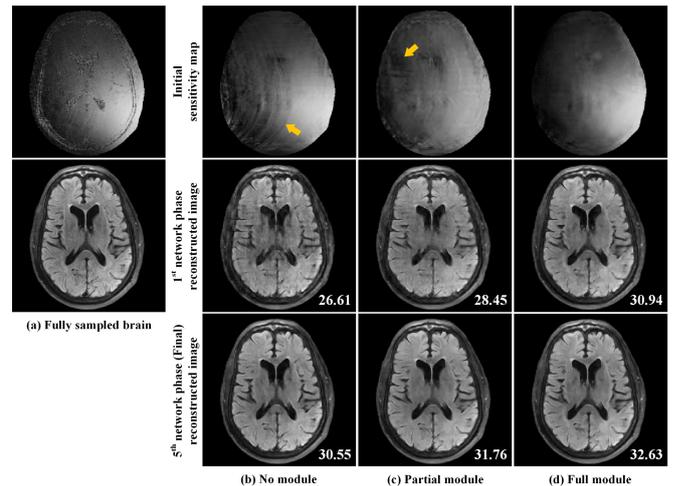

**Fig. 5.** Representative brain images reconstructed by networks with different initialization modules. (a) is the ground truth sensitivity map of $11^{th}$ coil and fully sampled image. (b)-(d) are initial sensitivity maps, reconstructed images of $1^{st}$ network phase, and reconstructed images of $5^{th}$ network phase (Final), from top to bottom. Note: The 1D Cartesian undersampling pattern with AF=8 is used. PSNR(dB) is listed for each reconstruction. The obvious artifacts are marked with yellow arrows.

Fig. 5 shows that both no and partial modules provide inferior sensitivity maps with obvious artifacts, while the proposed full module has better initial maps and directly leads to a much better reconstruction image in the 1st network phase (i.e., outperforms others by about 2.49~4.33 dB in PSNR). The gains continue until the end, resulting in the best reconstruction of the network with the full module. Consistent quantitative results of more slices are summarized in Table I.

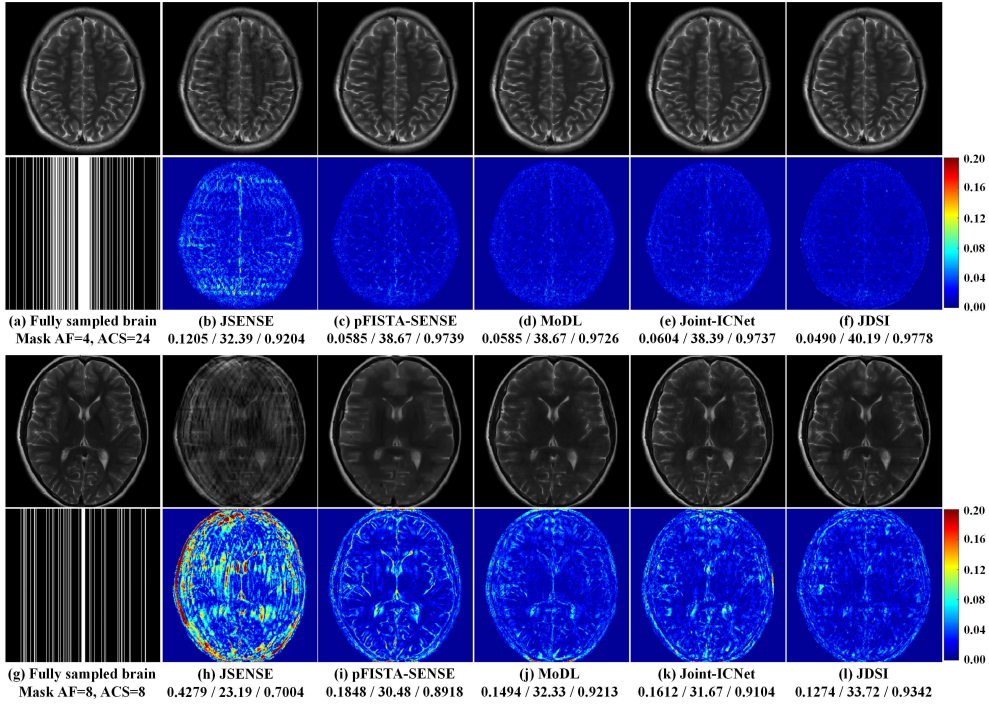

**Fig. 6.** Calibration-based reconstruction results of the T2 weighted brain dataset using different methods. (a) (or (g)) is the fully sampled image and the 1D Cartesian undersampling pattern with AF=4 (or AF=8). (b)-(f) (or (h)-(l)) are reconstructed images and the corresponding error maps, from top to bottom. Note: RLNE/PSNR(dB)/SSIM are listed for each reconstruction.

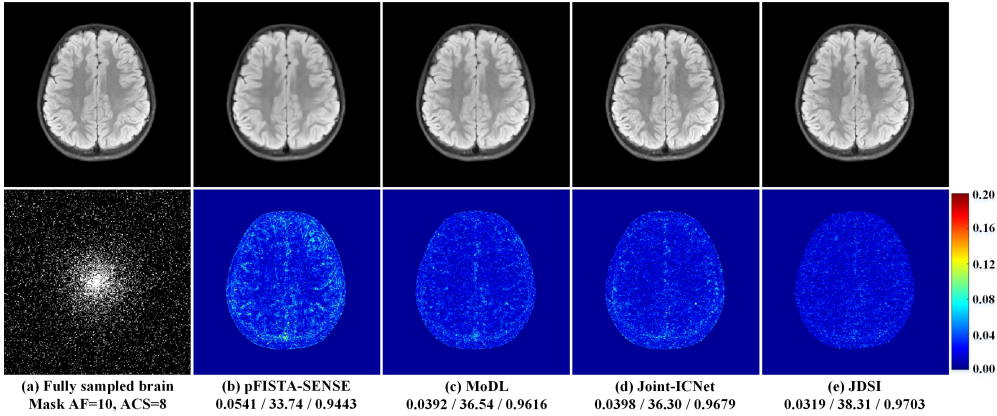

**Fig. 7.** Calibration-based reconstruction results of the T2 FLAIR brain dataset using different methods. (a) is the fully sampled image and the 2D random undersampling pattern with AF=10. (b)-(e) are reconstructed images and the corresponding error maps, from top to bottom. Note: RLNE/PSNR(dB)/SSIM are listed for each reconstruction.

These results demonstrate that the proposed module can generate the best initial sensitivity maps. It helps the network reconstruct a proper image from the beginning and benefits the final reconstructions indirectly.

*2) Joint loss function*

Then, we evaluated the importance of the proposed joint loss function. It not only minimizes the error of images, but also explicitly constrains sensitivity maps in the loss function.

Considering the commonly used image loss $\mathcal{L}_{image}$ as the basic loss function. In Table II, we find that the reconstruction performance can be improved by adding an explicit constraint $\mathcal{L}_{sensitivity}$ on sensitivity maps to minimize the error of them (i.e., joint loss function in Eq. (10)). Take PSNR as an example, the improvement is 0.36 dB.

These results imply that, involving the sensitivity loss in training is an effective strategy to improve reconstructions.

*3) Joint network architecture*

Finally, we evaluated the importance of the proposed joint network architecture, which simultaneously updates sensitivity maps during image reconstruction.

Consider a network using fixed sensitivity maps as the basic network architecture. In Table III, it is found that by further inserting the sensitivity estimation module as shown in Eq. (6), the joint network can dramatically boost the quality of reconstructed images, e.g., PSNR improves by 2.20 dB.

These results show that simultaneous sensitivity update and image reconstruction as proposed is a major source of image quality improvement.

**TABLE I**
RLNE (×10$^{-2}$)/PSNR (dB)/SSIM (×10$^{-2}$) OF BRAIN RECONSTRUCTIONS USING NETWORKS WITH DIFFERENT INITIALIZATION MODULES [MEAN±STD].

| 1D Cartesian undersampling pattern (AF=8, ACS=8) | | | | |
|---|---|---|---|---|
| Dataset | Method | RLNE | PSNR | SSIM |
| T2 FLAIR | No module | 8.50±1.71 | 32.84±2.18 | 92.76±1.89 |
| | Partial module | 8.07±1.68 | 33.31±2.11 | 93.16±1.76 |
| | Full module (JDSI) | **7.22±1.53** | **34.28±2.18** | **93.86±1.75** |

Note: The means and standard deviations are computed over all test slices, respectively. The lowest RLNE, highest PSNR and SSIM values are bold faced.

**TABLE II**
RLNE (×10$^{-2}$)/PSNR (dB)/SSIM (×10$^{-2}$) OF BRAIN RECONSTRUCTIONS USING NETWORKS TRAINED BY DIFFERENT LOSS FUNCTIONS [MEAN±STD].

| 1D Cartesian undersampling pattern (AF=8, ACS=8) | | | | |
|---|---|---|---|---|
| Dataset | Method | RLNE | PSNR | SSIM |
| T2 FLAIR | Image loss | 7.51±1.53 | 33.92±2.00 | 93.71±1.62 |
| | Joint loss (JDSI) | **7.22±1.53** | **34.28±2.18** | **93.86±1.75** |

Note: The means and standard deviations are computed over all test slices, respectively. The lowest RLNE, highest PSNR and SSIM values are bold faced. The network architectures of these two are consistent with JDSI, except for the loss function.

**TABLE III**
RLNE (×10$^{-2}$)/PSNR (dB)/SSIM (×10$^{-2}$) OF BRAIN RECONSTRUCTIONS USING DIFFERENT NETWORK ARCHITECTURES [MEAN±STD].

| 1D Cartesian undersampling pattern (AF=8, ACS=8) | | | | |
|---|---|---|---|---|
| Dataset | Method | RLNE | PSNR | SSIM |
| T2 FLAIR | Basic network | 9.23±1.62 | 32.08±2.03 | 92.16±1.86 |
| | Joint network (JDSI) | **7.22±1.53** | **34.28±2.18** | **93.86±1.75** |

Note: The means and standard deviations are computed over all test slices, respectively. The lowest RLNE, highest PSNR and SSIM values are bold faced.

### C. Comparison with State-of-the-art Methods

#### 1) Calibration-based reconstruction

To validate the advantages of the proposed JDSI, we first compared it with the state-of-the-art methods under common calibration-based undersampling patterns. These patterns have fully sampled central ACS regions for sensitivity encoding or k-space kernel learning, to model the relationship between multiple coils.

For the comparison study, the conventional parallel imaging method JSENSE [12] was used as a reconstruction baseline (Only supports 1D undersampling). We also compared the state-of-the-art sparse reconstruction method pFISTA-SENSE [11], and two calibration-based deep learning methods including MoDL [25] and Joint-ICNet [32]. JSENSE is a classical joint sensitivity estimation and image reconstruction method. pFISTA-SENSE and MoDL need pre-estimate sensitivity maps using ESPIRiT [6]. Joint-ICNet is a joint method but updates sensitivity maps using a conventional linear operation. Parameters of all conventional methods were optimized to obtain the lowest RLNE. The network phase of Joint-ICNet was set as 5 due to GPU memory limitation (16 GB). Other deep learning methods were executed according to typical settings mentioned by the authors.

For 1D undersampling, Figs. 6(a)-(f) show that, when the acceleration factor is low (AF=4), except JSENSE has obvious

**TABLE IV**
RLNE (×10$^{-2}$)/PSNR (dB)/SSIM (×10$^{-2}$) OF BRAIN RECONSTRUCTIONS UNDER CALIBRATION-BASED UNDERSAMPLING PATTERNS [MEAN±STD].

| 1D Cartesian undersampling pattern (AF=4, ACS=24) | | | | |
|---|---|---|---|---|
| Dataset | Method | RLNE | PSNR | SSIM |
| T2 weighted | JSENSE | 12.62±1.58 | 30.97±1.14 | 90.25±1.92 |
| | pFISTA-SENSE | 6.53±1.48 | 36.93±1.39 | 96.28±1.27 |
| | MoDL | 6.52±1.46 | 36.82±1.42 | 96.29±0.97 |
| | Joint-ICNet | 7.20±1.47 | 35.93±1.37 | 95.93±1.46 |
| | JDSI | **6.07±1.80** | **37.56±1.86** | **96.39±1.71** |
| T2 FLAIR | JSENSE | 9.95±1.52 | 31.39±1.98 | 89.40±2.11 |
| | pFISTA-SENSE | 6.53±1.28 | 35.12±1.75 | 94.78±1.25 |
| | MoDL | 5.34±1.06 | 36.88±1.81 | 96.53±0.83 |
| | Joint-ICNet | 4.68±0.80 | 37.97±1.97 | 96.65±0.88 |
| | JDSI | **3.99±0.79** | **39.38±2.19** | **97.36±0.82** |
| 1D Cartesian undersampling pattern (AF=8, ACS=8) | | | | |
| T2 weighted | JSENSE | 36.85±4.45 | 21.66±1.51 | 73.39±4.36 |
| | pFISTA-SENSE | 18.07±4.13 | 27.96±1.79 | 87.55±2.42 |
| | MoDL | 15.92±1.35 | 28.92±1.36 | 90.54±1.27 |
| | Joint-ICNet | 16.49±2.01 | 28.64±1.82 | 89.33±2.41 |
| | JDSI | **13.52±2.25** | **30.42±1.82** | **91.43±2.19** |
| T2 FLAIR | JSENSE | 24.20±3.04 | 23.63±1.97 | 79.89±3.35 |
| | pFISTA-SENSE | 18.86±3.06 | 25.84±2.57 | 85.92±2.96 |
| | MoDL | 13.41±2.65 | 28.87±1.44 | 90.00±1.70 |
| | Joint-ICNet | 8.08±1.42 | 33.23±2.19 | 93.73±1.84 |
| | JDSI | **7.22±1.53** | **34.28±2.18** | **93.86±1.75** |
| 2D random undersampling pattern (AF=10, ACS=8) | | | | |
| T2 FLAIR | pFISTA-SENSE | 8.41±1.49 | 32.89±1.56 | 92.45±1.58 |
| | MoDL | 6.15±1.08 | 35.61±1.58 | 94.99±0.89 |
| | Joint-ICNet | 5.93±0.98 | 35.90±2.10 | 95.66±1.39 |
| | JDSI | **5.05±0.93** | **37.34±2.05** | **95.84±1.18** |

Note: The means and standard deviations are computed over all test data. The lowest RLNE, highest PSNR and SSIM values are bold faced.

artifacts, other compared methods provide the images with nice artifacts suppression while JDSI achieves the best performance visually and quantitatively. However, when the acquisition is highly accelerated (AF=8), all compared methods own dramatic degradation and yield results exhibiting obvious artifacts, whereas JDSI has the smallest reconstruction error and outperforms them by about 1.39~10.53 dB in PSNR (Figs. 6(g)-(l)). For 2D undersampling, Figs. 7(a)-(e) show that JDSI provides the best reconstructions, while compared methods have edge artifacts and details blurring. Consistent quantitative results of more datasets are summarized in Table IV.

These results demonstrate that, for calibration-based reconstruction, JDSI can achieve the best ability of the artifacts suppression and details preservation, under different tested undersampling patterns and acceleration factors.

#### 2) Calibrationless reconstruction

In general, the need to obtain ACS signals prolongs the overall data acquisition time and sometimes increases the susceptibility to subject motion [14, 15]. Therefore, calibrationless reconstruction has become a research hotspot due to it does not rely on ACS signals (i.e., no ACS) [14, 15, 23, 24]. Different from other sensitivity-based methods, our proposed JDSI estimates and updates the sensitivity maps from the whole k-space instead of the central ACS region, leading to a potential for successful calibrationless reconstruction. Here, we compared the proposed JDSI with the state-of-the-art

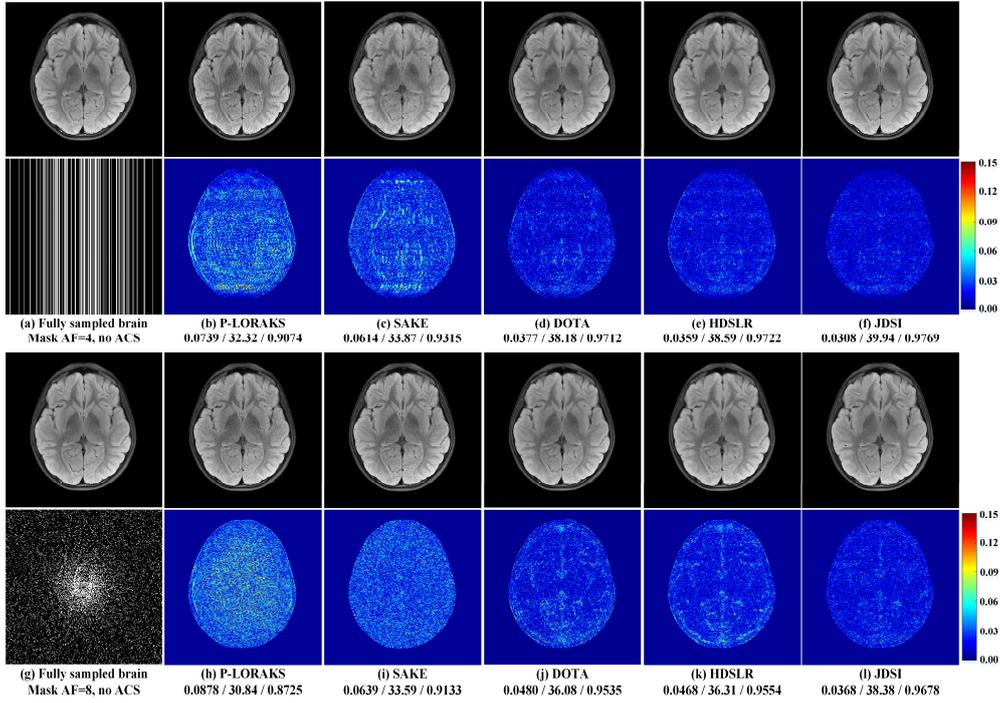

**Fig. 8.** Calibrationless reconstruction results of the T2 FLAIR brain dataset using different methods. (a) (or (g)) is the fully sampled image and the corresponding undersampling pattern. (b)-(f) (or (h)-(l)) are reconstructed images and the corresponding error maps, from top to bottom. Note: RLNE/PSNR(dB)/SSIM are listed for each reconstruction.

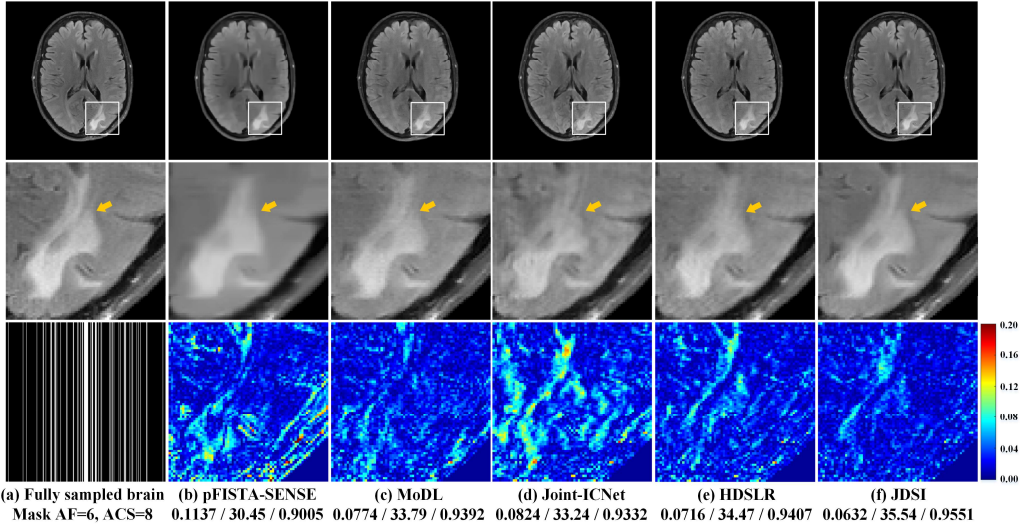

**Fig. 9.** Reconstruction results of the T2 FLAIR brain dataset with lesion using different methods. (a) is the fully sampled image, magnified details, and the corresponding undersampling pattern. (b)-(f) are reconstructed images, magnified details, and the corresponding error maps, from top to bottom. Note: The white matter lesions are marked with yellow arrows. RLNE/PSNR(dB)/SSIM are listed for each reconstruction.

calibrationless methods.

For the comparison study, two representative low-rank modeling methods P-LORAKS [15] and SAKE [14], and two calibrationless deep learning methods DOTA [22] and HDSLR [23] were compared. Parameters of all conventional methods were optimized to obtain the lowest RLNE. The deep learning method was executed according to the typical setting mentioned by the authors.

Calibrationless reconstruction is more challenging than the former calibration-based reconstruction. As expected, the results shown in Fig. 8 and Table V are worse than those in the patterns with ACS signals. Interestingly, consistent with the previous results, the proposed JDSI still outperforms all compared methods visually and quantitatively in both 1D and 2D undersampling. DOTA and HDSLR perform well on the 1D pattern but yields obvious artifacts and image blurring on the 2D pattern, while other two low-rank methods cannot achieve satisfactory images on all tested undersampling patterns.

These results demonstrate that, for calibrationless reconstruction without ACS, JDSI can still obtain the lowest reconstruction error and highest structural similarity under different tested undersampling patterns and acceleration factors, again suggesting its effectiveness and flexibility.

## D. Robustness to Patients

In general, the training datasets are mainly composed of healthy volunteers because they are relatively easy to conduct large-scale acquisition. However, the data of patients are often different from that of healthy subjects, mainly reflected in the distribution and intensity of pathological tissues. These differences between the training and target data may lead to the performance drop of deep learning methods. Thus, exploring the robustness of the proposed method to patients is essential for clinical applications.

Here, we used a brain dataset of healthy subjects to train all deep learning methods, and then to reconstruct a brain image with white matter lesion. The patient data were obtained according to the information provided in fastMRI+ [40] which were annotated as non-specific white matter lesion.

Figs. 9(a) and (f) show that the proposed JDSI recovered the pathological tissues (Marked with yellow arrows) most like the fully sampled image. Whereas other compared methods do not recover the lesion well, resulting in over-smooth (Fig. 9(b)), structure loss (Fig. 9(c)), intensity loss (Fig. 9(d)), and residual artifacts (Fig. 9(e)).

The observation demonstrates that the proposed JDSI owns some robustness to patients.

**TABLE V**
RLNE (×10$^{-2}$)/PSNR (dB)/SSIM (×10$^{-2}$) OF BRAIN RECONSTRUCTIONS UNDER CALIBRATIONLESS UNDERSAMPLING PATTERNS [MEAN±STD].

| Dataset | Method | RLNE | PSNR | SSIM |
|---|---|---|---|---|
| \multicolumn{5}{c}{1D Cartesian undersampling pattern (AF=4, no ACS)} | | | | |
| T2 FLAIR | P-LORAKS | 9.22±1.64 | 32.08±2.24 | 91.38±1.45 |
| | SAKE | 9.02±1.90 | 32.33±2.17 | 91.40±2.32 |
| | DOTA | 5.02±0.91 | 37.39±2.33 | 96.72±0.93 |
| | HDSLR | 4.95±0.90 | 37.51±2.07 | 96.12±1.08 |
| | JDSI | **4.33±0.86** | **38.69±2.16** | **97.15±0.83** |
| \multicolumn{5}{c}{2D random undersampling pattern (AF=8, no ACS)} | | | | |
| T2 FLAIR | P-LORAKS | 10.99±1.53 | 30.50±2.14 | 87.79±2.04 |
| | SAKE | 9.47±1.86 | 31.89±2.30 | 88.83±2.94 |
| | DOTA | 6.38±1.15 | 35.30±2.17 | 94.74±1.35 |
| | HDSLR | 5.94±0.97 | 35.89±1.92 | 95.14±1.20 |
| | JDSI | **4.95±0.90** | **37.51±2.07** | **96.12±1.08** |

Note: The means and standard deviations are computed over all test slices, respectively. The lowest RLNE, highest PSNR and SSIM values are bold faced.

## E. Robustness to Different ACS Lines

The ACS is crucial for accelerated MRI reconstruction, once the number of ACS lines is limited, the pre-estimated sensitivity maps become inaccurate, leading to the degradation of reconstruction [12, 14]. Thus, refining sensitivity maps during the reconstruction may be important.

Here, to validate the robustness of our joint network to different ACS lines, we compared the proposed JDSI with other two networks which use pre-estimated sensitivity maps from JSENSE [12] and ESPIRiT [6], respectively. The network architectures of these two are consistent with JDSI, except for removing the sensitivity-update-related modules.

Fig. 10 shows that, for the tested number of ACS lines, the proposed JDSI consistently outperforms other two methods in terms of three evaluation criteria. Moreover, with a decrease of ACS lines (<16 lines), the accuracy of two pre-estimated sensitivity maps decreases, resulting in significant evaluation criteria degradations. By updating the sensitivity maps in the network, these criteria of the proposed JDSI are only slightly reduced when the ACS lines are reduced. Thus, updating the sensitivity maps provides robustness to ACS lines.

Representative images in Figs. 11(a)-(d) show that, when ACS lines are enough (ACS=24), all three methods can yield good sensitivity estimations and suppress artifacts well. The sensitivity map of JDSI is closest to ground truth due to sensitivity refinement. However, when the ACS lines are limited (Figs. 11(e)-(h)), JSENSE and ESPIRiT provide inferior sensitivity maps, resulting in obvious artifacts in reconstructed images, whereas our JDSI provides consistently superior sensitivity estimation with nice image artifacts suppression.

It demonstrates the importance of faithful sensitivity estimation for high-quality image reconstructions. Our joint network overcomes the limitation of conventional sensitivity estimation strategies under limited ACS lines, providing high-quality and robust reconstructions.

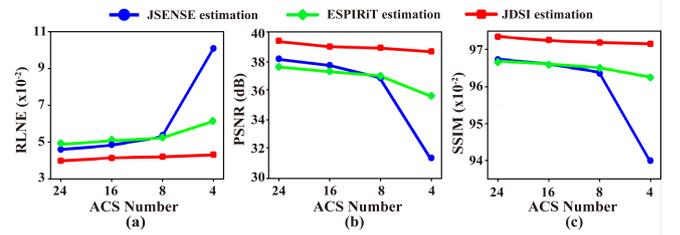

**Fig. 10.** Quantitative comparison of reconstructions using different methods with different number of ACS lines of a brain dataset. (a)-(c) are the means of RLNE, PSNR, SSIM of the reconstructions, respectively. Note: The 1D Cartesian undersampling pattern with AF=4 is used. The difference between each mask is the number of ACS lines. The means are computed over all test slices.

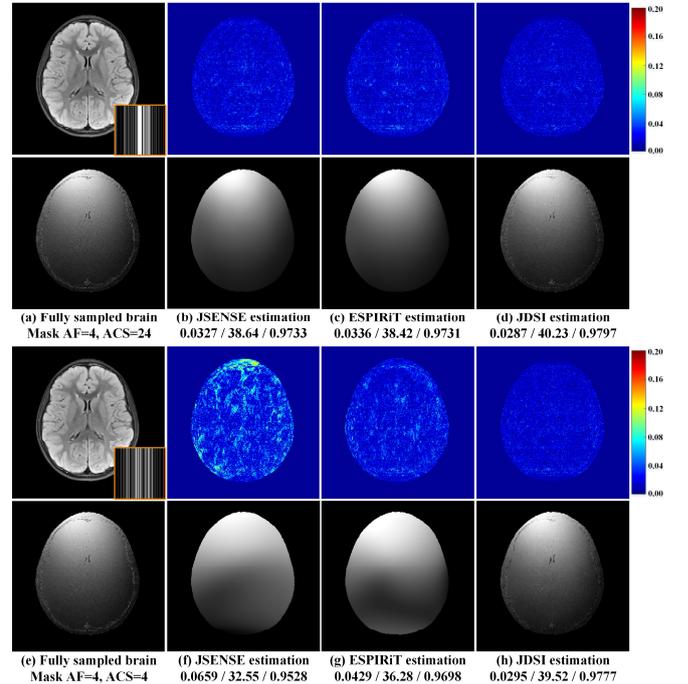

**Fig. 11.** Representative reconstructed brain images with and without enough ACS lines. (a) (or (e)) is the fully sampled image, the corresponding undersampling pattern, and the ground truth sensitivity map of 14$^{th}$ coil. (b)-(d) (or (f)-(h)) are reconstruction errors and estimated sensitivity maps, from top to bottom. Note: The 1D Cartesian undersampling patterns are used. RLNE/PSNR(dB)/SSIM are listed for each reconstruction.

## F. Reader Study

Since objective criteria (i.e., RLNE, PSNR, and SSIM) might not reflect image quality in terms of diagnostically important features, we further invited three radiologists (with 12, 29, and 30 years of clinical experience) to independently evaluate the reconstruction images. They were blind to the reconstruction methods. To avoid rater bias, the order of reconstructed images was presented randomly for each slice. Three clinical-concerned criteria including signal-to-noise ratio, artifacts, and overall image quality were used (Table VI). The reader study was conducted through our cloud platform [21, 41, 42] at https://csrc.xmu.edu.cn/CloudBrain.html.

The images used for the reader study were T2 FLAIR images reconstructed under the 1D Cartesian undersampling pattern with AF=6. They included 10 healthy subjects and 10 patients. The patient data were obtained according to the information provided in fastMRI+ [40] which were annotated as non-specific white matter lesion. To reduce the burden of radiologists' scoring, each subject contained about central three slices which have sufficient tissue information.

Table VII and Fig. 12 show that the proposed JDSI obtains highest mean scores. It is the only one which all three criteria are slightly over 4, indicating the reconstructed images are suitable for diagnosis. These results are consistent with the superiority of JDSI on the above-mentioned objective criteria. The scores of other compared methods mean their images are good for diagnosis. Besides, the differences between JDSI and other compared methods are statistically significant according to all P-values of the Wilcoxon signed rank test < 0.001.

In summary, the image quality improvements obtained by the proposed JDSI is significant and the overall quality steps into the excellent level for clinical diagnosis.

**TABLE VI**
5-POINT LIKERT SCALE SCORING CRITERIA FOR THE READER STUDY.

| Score | Signal-to-noise ratio | Artifacts | Overall image quality |
|---|---|---|---|
| 5 | Excellent | None | Excellent |
| 4 | Good | Minimal | Good |
| 3 | Fair | Acceptable | Adequate |
| 2 | Low | Significant | Poor |
| 1 | Poor | Massive | Non-diagnostic |

Note: The score of each criterion has a range from 0 to 5 with precision of 0.1.

**TABLE VII**
THE SCORES OF THE READER STUDY [MEAN±STD].

| Method | Signal-to-noise ratio | Artifacts | Overall image quality |
|---|---|---|---|
| MoDL | 3.59±0.30 | 3.65±0.44 | 3.64±0.29 |
| Joint-ICNet | 3.72±0.22 | 3.59±0.27 | 3.70±0.20 |
| HDSLR | 3.51±0.15 | 3.52±0.15 | 3.53±0.15 |
| JDSI | **4.01±0.19** | **4.01±0.18** | **4.02±0.18** |

Note: Images from 10 healthy subjects and 10 patients are involved in the reader study. The means and standard deviations are computed over all images, respectively. The highest scores are bold faced.

## G. Comparison with the State-of-the-art E2E-VarNet

Here, we further compared another state-of-the-art deep learning methods, called E2E-VarNet [31]. It employs a U-Net as the sensitivity estimator at the beginning of the network to avoid time-consuming ESPIRiT pre-estimation [6], and the end-to-end training manner makes it superior to some methods that rely on pre-estimated sensitivity maps. However, E2E-VarNet only uses the ACS region for sensitivity estimation and loses high-frequency information, while ignores the importance of updating sensitivity maps during the reconstruction process. These problems may degrade final reconstructions, especially when the AF is high.

Differently, our proposed JDSI uses the whole k-space rather than only the ACS region to capture more high-frequency information, while integrates a customized sensitivity estimation module to update sensitivity maps and make them further approximate the ground truth sensitivity maps. Thus, JDSI can simultaneously achieve more faithful sensitivity estimation and high-quality image reconstruction.

Figs. 13(a)-(c) show that, when the AF is low, both two methods can yield good sensitivity estimations and image reconstructions. Besides, JDSI achieves better reconstruction both visually and quantitatively, since its sensitivity map is much closer to ground truth. However, for a higher AF in Figs. 13(d)-(f), E2E-VarNet provides inferior sensitivity map, resulting in image artifacts and the lesion structure loss, whereas our JDSI provides consistently superior sensitivity estimation with nice image artifacts suppression and lesion details preservation.

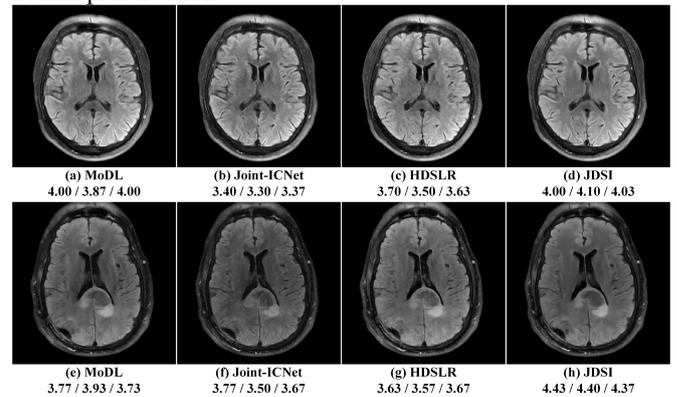

**Fig. 12.** Representative images for the reader study. (a)-(d) are reconstructed images from a healthy subject. (e)-(h) are reconstructed images from a patient. Note: Each image lists the mean scores of three radiologists in the signal-to-noise ratio/artifacts/overall image quality.

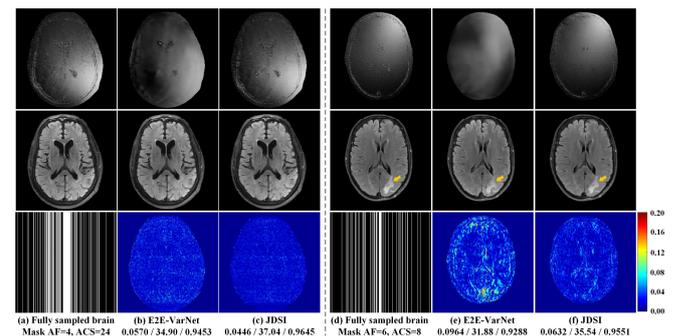

**Fig. 13.** Reconstruction results of the T2 FLAIR brain dataset (left: healthy subject, right: patient) using different methods. (a) (or (d)) is the ground truth sensitivity map of $11^{th}$ (or $1^{st}$) coil, the fully sampled image, and the corresponding undersampling pattern. (b)-(c) (or (e)-(f)) are estimated sensitivity maps, reconstructed images, and the corresponding error maps, from top to bottom. Note: The white matter lesions are marked with yellow arrows. RLNE/PSNR(dB)/SSIM are listed for each reconstruction.

## V. Discussions

In this work, we first clearly demonstrate the key role of a faithful coil sensitivity estimation with high-frequency information in MRI reconstruction. Following this idea, we then elaborately design network modules that estimate sensitivity maps using all measured k-space and explicitly constrain these maps in the loss function to obtain better sensitivity maps. Integrating them into a deep image reconstruction network, we successfully observe the mutual promotion of sensitivity estimation and image reconstruction, which gradually leads to a final high-quality image. More importantly, the clear design motivation and network behavior also make our further optimization traceable.

An impressive advantage of the proposed JDSI is its superior performance in both healthy subjects and patients when imaging acceleration is high and/or autocalibration signals are limited. In particular, JDSI can clearly reconstruct the margin and contrast of lesions. This is conductive to better monitoring of subtle changes in lesion structures during follow-up to determine whether the condition is improving (lesion absorption) or deteriorating (lesion expansion). Besides, its reliability is also supported by three experienced radiologists from a clinical diagnostic perspective.

In the future, we anticipate that the following aspects are worth exploring: 1) Test its usability of other anatomies (e.g., knee, spine, and cardiac) and attempt to utilize emerging networks [43, 44] for further optimization or adaptation. 2) Extend the idea of deep coil sensitivity estimation to other MRI applications that rely on accurate coil correlation modeling and require high resolution and acceleration, such as dynamic and quantitative MRI.

## VI. Conclusion

In this work, we present a Joint Deep Sensitivity estimation and Image reconstruction network (JDSI) for accelerated magnetic resonance imaging (MRI). Our network simultaneously achieves the faithful estimation of sensitivity maps and the reconstruction of high-quality images.

Extensive results on *in vivo* datasets demonstrate that, for both calibration-based and calibrationless reconstruction, the proposed JDSI provides improved and more robust reconstruction performance than state-of-the-art methods. Besides, the reliability of our JDSI in the diagnostic perspective is validated by three experienced radiologists even under highly accelerated scanning.

In summary, a faithful deep sensitivity estimation is expected to be beneficial for clinical MRI applications, which need high imaging accelerations and/or have limited acquisition of autocalibration signals to improve patient comfort and reduce motion-induced artifacts.

## Acknowledgment

The authors thank Yirong Zhou and Jiayu Li for supporting the reader study on the Cloud Brain Imaging platform; Xinlin Zhang and Dan Ruan for the helpful discussions; Drs. Michael Lustig, Mathews Jacob, Dosik Hwang, Justin P. Haldar, and Anuroop Sriram for sharing their codes online; the editors and reviewers for the valuable comments.